\documentclass[twocolumn,floatfix,showpacs,aps,prl]{revtex4}
\usepackage{graphicx}
\usepackage{epsfig}
 \begin{document}
\setlength{\unitlength}{1mm}
\bibliographystyle{unsrt} 
\title{  Linear Stark Shifts to Measure  the Fr Weak Nuclear Charge with Small Atom Samples}
\author{Marie-Anne Bouchiat  }
\affiliation{ Laboratoire Kastler Brossel, D\'epartement de Physique de l'Ecole
Normale Sup\'erieure,\\ 24 Rue Lhomond, F-75231  Paris  Cedex 05, France }
\date {November 5, 2007}
 \newcommand \be {\begin{equation}}
\newcommand \ee {\end{equation}}
 \newcommand \bea {\begin{eqnarray}}
\newcommand \eea {\end{eqnarray}}
\newcommand \nn \nonumber 
\def \(({\left(}
\def \)){\right)}
 \def \vs{{\mathbf{s}}}    
 \def \vI{{\mathbf{I}}}
 \def \vr{{\mathbf{r}}}
 \def \vF{{\mathbf{F}}}
 \def \vr{{\mathbf{r}}}
\def \vE{{\mathbf{E}}}
\def \vB{{\mathbf{B}}}
\def \ve{{\mathbf{e}}}
\def \vk{{\mathbf{k}}}
\def \vD{{\mathbf{D}}}
\def \vd{{\mathbf{d}}}
\def\bra{\langle}
\def\ket{\rangle}

\begin{abstract}We study the chirality of ground-state alkali atoms in $\vE$ and $\vB$ fields, dressed with a 
circularly-polarized laser beam detuned from an E-field-assisted forbidden transition, such as $7S-8S$ in Fr.
We predict a parity violating energy shift of their sublevels, linear in $\vE$, and the weak nuclear charge
$Q_W$.  A dressing beam of 10 kW/cm$^2$ at 506 nm produces a shift of     
$\sim100~\mu$Hz at E=100~V/cm, B $\gtrsim$30~mG. It    
should be observable with $\sim 10^4$ Fr atoms confined in an optical dipole
trap. We discuss optimal conditions, parameter reversals and a calibration procedure to measure 
$Q_W$.        
 \end{abstract}
\maketitle
 Atomic physics experiments in cesium have succeeded in determining  
the parity violating (PV) electric-dipole $6S-7S$ transition amplitude,
$E_1^{pv}$ as small as $0.8 \times 10^{-11} $ e$a_0$, with a  
precision progressively \cite{bou74} reaching 0.35 $\%$, once both experimental \cite{ben99} and absolute 
calibration \cite{bou88} uncertainties are included. Thanks to the {\it tour de force} made by
atomic theorists \cite{fla04}, the weak Cs nuclear charge,  $Q_W$, extracted
from the result leads to a unique low-energy test of the standard
electroweak theory. Strong incentives for still improving its 
precision, {\it e.g.} for reducing the limits on the mass of an eventual additional, light or heavy, 
neutral boson \cite{ZAPV}, motivate the present work. 

One possibility is to take full advantage
of the PV asymmetry amplification mechanism already  demonstrated in Cs \cite{gue03}. This occurs when 
the optical gain of a vapor relative to an allowed-transition frequency, is affected by the PV asymmetry and probed
using a  transmitted beam. The intensity of the beam grows exponentially along its path and, more
importantly, so does the PV asymmetry.  
At higher gains, a dramatic amplification of the PV asymmetry is expected for low 
probe intensities sufficient to
trigger superradiance on one $7S-6P$ line \cite{bou05}. 
This results in a largely improved  quantum noise limited
signal/noise ratio.

However, another approach frequently suggested deals with PV measurements
in radioactive francium, a very exciting, though extremely challenging
project \cite{franc}. In this atom, the heaviest of the alkalis, the
computed value of $E_1^{pv}$  for the  $7S-8S$ transition is 18 times larger
than in Cs $6S-7S$ \cite{fla04}. Several spectroscopic measurements have already been performed
on cold Fr atoms inside a magneto optical trap (MOT), but even the observation of this forbidden    
transition still looks difficult. The main problem is that, contrary to Cs, the number of available
atoms is small: no more than $10^4$ up to now. Thus, transpositions from Cs to
Fr of PV detection
schemes successful in the past \cite{san05}, would be doomed to failure. All of them used at least $10^7$ and even 
up to $10^{12}$ atoms   
for detecting a forbidden-transition polarization-dependent
probability on a light beam intensity, observed either by fluorescence or transmission.
By contrast, frequency measurements based on atom interferometry have demonstrated unprecedented
accuracy and proved to be well adapted to cold samples of $10^4$ to $10^6$ atoms. Therefore, the existence of a
PV effect attainable by these methods could form the basis of a realistic Fr project. 
The problem is that an electric dipole moment violating P but not T cannot give rise to a frequency 
shift in a stationary atomic system submitted only  to homogeneous electric and magnetic fields, as stated by  
Sandars' theorem
\cite{lov}. The strategy is to find some way around this restriction. 
The light-shift proposal
for a single Ba$^{+}$ ion \cite{for93} based on the radiation field gradient in a standing wave,  is made
difficult by the need for precise adjustment of the ion position in this field. For the Cs nuclear
anapole moment, we have suggested measuring the hyperfine (hf) linear Stark-shift of the ground state  dressed
by a circularly polarized laser beam detuned from the first resonance line \cite{mab07}.  
 In this case the measurement involves off-diagonal matrix elements of the PV electric dipole between hf ground 
states. No similar effect exists for the $Q_W$  electric dipole which has no such 
 matrix elements. In this Letter we show that, if the dressing
beam is close to resonance with respect to the Stark-field-assisted $nS-n'S$ transition, an 
interferometric measurement of $Q_W$ may also be achieved by  observing a linear Stark-shift. 

 We consider alkali atoms in their ground state, placed in crossed electric and magnetic dc fields, $\vE$
and $\vB$, which interact with a laser beam in a direction $\hat k$ orthogonal to both fields. Its angular   
frequency $\omega$, is detuned from the 
forbidden, E-field assisted, $nS - n^{\prime}S$ transition of frequency $\omega_0$ by an amount $\delta$.
Opposite parity states are mixed by the electric dipole coupling $V_E= - \vD \cdot \vE$. 
 A second mixing between $S_{1/2}$ and $P_{1/2}$ states results from the PV electron-nucleus interaction
$V_{pv}$, associated with the weak nuclear charge $Q_W$~\cite{bou74},   
$\bra n^{\prime}\, P_{1/2} \vert V_{pv}\vert n\,
S_{1/2}\ket = - i \,G_F \,Q_W  \bra n^{\prime}\, P_{1/2}\vert {\cal{R}} \vert n\,
S_{1/2}\ket   $.
The factor $i$ results  from its time reversal invariance, $G_F$ is the Fermi constant and 
$\cal{R}$ a short-range pseudoscalar operator mixing the $S_{1/2}$ and $P_{1/2}$ wave functions in the
vicinity of the nucleus.    
 
To describe the electric dipole amplitudes induced by the radiation field as a result of both mixings, it is convenient
to introduce the effective electric dipole operator which behaves as a matrix element in the radial atom coordinate
space and as an operator in the spin space,  
\begin{eqnarray}
\vD^{ef}(n,n^{\prime})&=& -\alpha(n,n^{\prime}) \vE  -i \, \beta(n,n^{\prime}) \vec \sigma \wedge \vE 
\nonumber\\ && - i \, {\rm Im} E_1^{pv}(n,n^{\prime}){\vec  \sigma} + M_1{\vec \sigma} \wedge \vk,   
\end{eqnarray}
where $\vec \sigma $ is the Pauli matrix for the electron spin operator. 
The final term describes the effective contribution of the parity conserving (PC) magnetic dipole amplitude
\cite{bou74,joh99}, in which, for both Cs and Fr,  
$M_1\approx  2 \times 10^4\,  \rm{Im}  E_1^{pv}$.

We use the quantized form of the radiation field, involving the creation and annihilation operators $ a
^\dagger $ and $a$.  The
circularly polarized laser beam is described by a normalized
$N$  photon state $ \vert N \ket $ with energy  density 
$\epsilon_0\,{\cal E  }^2 = N\, \hbar  \omega /V$ localized inside a volume $V$. The Hamiltonian 
$V_{rad} = -\sqrt {\frac {\hbar \omega}{2 \epsilon_0 V}} (a \; \ve(\xi) \cdot  \vD + h.c. )$ describes
the atom-field dipolar coupling with $  \ve (\xi) =\frac{1}{\sqrt{2}}\, ( {\ve
}_1 +i\,\xi \,{\ve }_2)$, $ \ve_i$  being two real orthogonal  unit vectors normal  to
the direction $ \hat k $  of the laser beam and $\xi = \pm1$ defining the helicity.
For the combined atom-field initial and final states, $\vert \tilde{i}\ket =\vert n\, \widetilde{S_{1/2}}
; F, m_{_F}\ket
\, \vert N \ket$ and  $\vert \tilde{f } \ket =\vert n^{\prime}\, \widetilde {S_{1/2}}; F^{\prime},
m^{\prime}_{_F}\ket \, \vert N-1 \ket  $, perturbed by
$ V_E + V_{pv}$, an effective dipolar coupling is described by the Hamiltonian $V^{ef}_{rad} = -\sqrt {\frac
{\hbar \omega}{2 \epsilon_0 V}} (a \; \ve(\xi) \cdot  \vD^{ef} + h.c. )$; 
$ F, m_{_F} $ are the hf and magnetic quantum numbers.     

Using this approach, the general expression of the $7S$ state light-shift arising from virtual absorption and
emission of close-detuned 
$7S_{1/2}-8S_{1/2}$ photons is   
\be 
\delta E =  \vert\bra \tilde{i}
\vert  V_{rad}^{ef} \vert \tilde{f} \ket\vert^2/ \hbar \delta . 
\ee
The fact that $V_{rad}^{ef}$ appears alone might imply that other significant contributions have been omitted.    
Indeed, in a rigorous perturbative treatment one should consider all fourth-order terms (linear in $V_{pv}, \, V_E,
\,$ and quadratic in $V_{rad}$). In fact, the only missing terms are strongly
``non-resonant'' for the present laser-detuning.
Only for a quasi-resonant dressing beam do they dominate, giving rise, then, to the
anapole shift evaluated previously \cite{mab07}.   
 
In the expression (2) we find all the terms making up the
operator $\vD^{ef}\cdot \ve (\xi)$  with their Rabi frequencies, the two scalar ($q=\alpha$)
and vector ($q=\beta$) Stark-induced ones, $\Omega_{ind}^q =
 q E  \, {\cal{E}}/ \hbar \sqrt{2}$, in addition to the tiny parity violating one, $\Omega_{pv}= {\rm Im}
E_1^{pv}{\cal{E}}/ \hbar \sqrt{2}$ and the $2 \times 10^4$ times larger magnetic dipole one, 
$\Omega_{M_1}= M_1 \mu_B {\cal{E}}/cea_0$.  The squared terms are similar in form to the light-shifts  
of the allowed transition, except for their noticeably smaller size. However, the energy shift $\delta
E_{F,m_F}$ we are looking for, results from an interference between the dominant Stark-induced
amplitude $\Omega_{ind}^{\alpha}$, and amplitudes $ \Omega_{pv}$, $\Omega_{M_1}$,  
\be 
\delta E_{F,m_F}= 2 \frac{ \hbar \Omega_{ind}^{\alpha}(\Omega_{M_1}+\xi \Omega_{pv})}{\delta_{_F}}
 \bra F,m_{_F}\vert \hat  E \wedge \hat k \cdot  
  \vec \sigma \vert F,m_{_F}\ket ,  
\ee
where $\delta_{_F}$ denotes the laser detuning with respect to the $nS_F \rightarrow n^{\prime}S_F$, ($\Delta F =
0$) hf line. It is convenient to introduce   the photon angular momentum $\xi \vk= i \, (\ve(\xi) \wedge
\ve^{*}(\xi)) \hbar$.
 Because, as justified later on, the direction $\hat B$ of the magnetic field (supposed weak,  
$\mu_B B \ll \Delta W$) still defines 
the quantization axis in the dressed state, we replace the
 matrix element in Eq. 3 by $g_{_F} m_{_F}$ with $g_{_F}= 2(F-I)/(I+1/2)$. The pseudoscalar $\chi = \hat  E
\wedge \xi \hat k \cdot \hat B$ is the mark of parity violation affecting the $E_1^{pv}$ contribution. From Eq.(3),
we derive the linear Stark frequency shifts $\Delta \nu_{hf}$ and $\Delta \nu_{Ze}$, for the 
$F^{\prime}=F-1 \rightarrow F, \Delta m_{_F}=0  $, hf transitions and 
the Zeeman ones $F,m_{_F} \rightarrow F, m_{_F} \pm 1$,  
\begin{eqnarray} 
&&\hspace{-5mm}\Delta \nu_{hf}= \frac{ 2 \Omega_{ind}^{\alpha}(\Omega_{M_1} + \xi 
\Omega_{pv})}{2\pi \delta_{_F}} 
 ( 1 + R_{_F}) \frac{m_{_F} \kappa_{_F}}{I+1/2} 
\,\, \hat E \wedge  \hat k \cdot \hat B , \nonumber \\  
&&\hspace{-5mm} \Delta \nu_{Ze} = 2 \Omega_{ind}^{\alpha}\frac{(\Omega_{M_1}+ \xi 
\Omega_{pv}) }{2 \pi \delta_{_F}} \kappa_{_F} \,\, \hat E \wedge  \hat k \cdot \hat B. 
 \end{eqnarray}
The quantity $R_{_F}= (\delta_{_F} /\delta_{_{F^{\prime}}})( \kappa_{_{F^{\prime}}}/\kappa	_{_F})$ involves
the detunings with respect to both $\Delta F=0$ transitions and 
 the factors $\kappa_{_F} = 1+
g_{_F} \beta/ \alpha$ and $\kappa_{_{F^{\prime}}}$, close to 1, which account for the contributions arising from
$\Omega_{ind}^{\beta}$.

In order to compensate for the weakness of the transition, one obvious solution is to strengthen the radiation
field.  This could be achieved by focusing the dressing beam on a small atom sample 
(0.01 to 1)mm$^3$ inside a radiative trap. However, two pitfalls have to be avoided: two-photon
ionization and radiative instability.  The former puts an upper limit on the magnitude of ${\cal{E}}$, while the
latter places a  constraint on the ratio $\Omega_{ind}^{\alpha}/\delta$, such that using large Stark fields
brings practically no benefit.  

 Multiphoton ionization processes, computed by quantum defect theory \cite{beb},    
 yields the frequency dependence of the two-photon ionization rate for all stable alkali atoms with
physical interpretation of peaks and valleys. Extending the results from Cs to Fr, using available spectroscopic data
\cite{franc}, we take as an estimate of the two-photon ionization rate for Fr at 506~nm     
$R  \,($s$^{-1})\approx 10^{-49}  \times $ (photon flux)$^2$ (nearly the same as for Cs at 539~nm). 
With a laser beam of 10 kW/cm$^2$ intensity ($2\times 10^{22}$ photons
s$^{-1}$/cm$^2$, and ${\cal{E}}=$ 2.2 kV/cm), we get conservatively  
$ R = 4\times 10^{-5}$~s$^{-1}$. No significant modification is expected for $E \leq 10$ kV/cm.
\squeezetable
\begin{table}[b]
\caption{The main atomic parameters of the Cs and Fr forbidden transitions defined in Eq.(1), from
\cite{bou88,fla04,ben99,joh99,san05,isold}, and the PV shift magnitudes (see Eq. 5 for the sign), for   
$\Delta m_{_F} =0, m_{_F} = I-1/2$ transitions, supposing $\vert \delta_{_F}/2 \pi \vert <
$~1GHz, $E$(V/cm)= 1.5 $\frac{\delta_F}{2 \pi}$ (MHz), ${\cal {E}  }$= 2.2 kV/cm. 
For different isotopes $\Delta \nu ^{pv}_{hf}$ varies as $(2I-1)/(2I+1)$ and $\Delta \nu ^{pv}_{Ze}$
as 1/(2I+1). }
\begin{center}
\begin{tabular}{p{3.4 cm} p{0.7cm} p{1.0cm}p{0.75cm}p{0.9cm}p{0.9cm}}
\hline\hline  $\small{\rm  ~~~Atomic~ transition}$&\small$_{ -\alpha/\beta}~$ &\scriptsize$
\vert E_1^{pv}\vert/\alpha$ 
\scriptsize 
$\rm mV/cm$
  &\scriptsize ${\rm M_1/\alpha }$ \scriptsize $ {\rm V/cm}$    &\small$~\Delta \nu ^{pv}_{hf}$ ~\scriptsize
$~~~(\mu\rm Hz)$ &\small $~\Delta \nu ^{pv}_{Ze}$
\scriptsize $ ~~(\mu\rm Hz)$ \\ 
\hline
 &&&&& ~~\\
\vspace{-5mm}$\scriptsize{^{133}{\rm Cs}, 6S-7S~(539~{\rm nm)}}$ & \small~~9.9 & \small ~~0.17   &
\small ~2.99  & \small~~~5.2 &\small ~~1.7 \\  
\vspace{-5 mm}$\scriptsize {I=7/2,\Delta W = 9.2~{\rm GHz}}$ & $ $ &  &     &  &   \\ 
\hline
&&&& & ~~\\
\vspace{-5mm}$\scriptsize{ ^{221}{\rm Fr},7S-8S~(506~{\rm nm)}}$ & \small $\simeq~7$ &\small~$~~2.2$  
&\small $\simeq~10 $  &\small$~~ 100 $& \small~~50  \\
\vspace{-5mm}$\scriptsize {I=5/2,\Delta W = 18.6~{\rm GHz}}$ & $ $ &  &     &  &   \\   
\hline \hline
\end{tabular}
\end{center}
\end{table}
On the other hand, the ground state levels acquire a finite lifetime characterized by
the  decay rate $\Gamma_{nS_F}=  \Gamma_{n'S} 
 (\Omega_{ind}^{\alpha}/\delta_{_F})^2 $. For precise interferometric measurements, the evolution of the atomic
system has to be observed over an interaction time, $\tau_i$, typically 1~s.  This means that the dressing intensity
has to be kept low enough to satisfy the ``stability condition'', $\Gamma _{nS_F}^{~~} \tau_i
\leq 1$, over this period. Hereafter, we shall assume this limit is just reached for the most tightly coupled hf
state, F.  This imposes the size of the ratio    
$\vert \Omega_{ind}^{\alpha}/\delta_{_F}\vert$ = $(\Gamma_{n'S}^{~~} \, \tau_i)^{-1/2}$, {\it i.e.} 2.2 $\times
10^{-4}$  for Cs ($n'$=7), 2.5 $\times 10^{-4}$ for Fr ($n'$=8); the sign is that of $\delta_{_F}$,
(since $\alpha>0$, and $\cal{E}$ and $E$ also by convention). The PV ($\chi$-odd) frequency shift
is, then, completely determined: 
\be
 \Delta \nu^{pv}_{hf} = (\Gamma_{n'S}^{~~} \tau_i)^{-1/2} \frac{\Omega^{pv}}{2\pi} 
( 1+ R_{_F})\, sign(\delta_{_F}) \frac{2m_{_F} \kappa_{_F}}{(I+1/2)}\, \chi,
\ee
with $0 <  R_{_F} < 1 $. The Stark field magitude $E$ is the single free parameter to determine
$\Omega^{\alpha}_{ind}$ once ${\cal{E}}$ is fixed. Since $E$ and $\delta_F$ are linked together by the stability
condition, they almost disappear from the result: $\Delta\nu^{pv}$ varies
only by a factor 2  from the low field range  ($\hbar \delta/\Delta W \ll 1 $), where the shift of a single hf state
dominates, to the high field one ($\hbar \delta/\Delta W\gg 1$), where both hf states participate.  In view of
the large E-field magnitudes such that $ \hbar \Omega_{ind}^{\alpha}(\Gamma_{n'S}
\tau_i)^{1/2} = \Delta W$ ($\sim$ 14 kV/cm for $^{133}$Cs, and 22 kV/cm for $^{221}$Fr), it looks more
realistic to envisage the measurement in the low field range, 20 to 1~000~V/cm, ($\delta_{_F}/2 \pi $=
13 to 650 MHz), much easier  to implement. Such a wide range leaves room for optimization of the
experiment. With a dressing beam intensity of 10~kW/cm$^2$, for Cs we obtain  
$\Omega^{pv}/2\pi= 23.5$~mHz and $\Delta 
\nu^{pv}_{hf} = 5.2 ~\mu$Hz for the
${3, \pm 3 \rightarrow 4, \pm 3} $  hf transition, while for Fr we expect  $\Omega^{pv}/2\pi\simeq 
0.42 $~Hz  leading to $\Delta  \nu^{pv}_{hf} \simeq 0.1~$mHz for the ${2, \pm 2 \rightarrow 3,
\pm 2}$ hf transition; in all cases $\Delta \nu^{pv}_{Ze}= \Delta  \nu^{pv}_{hf}/(I-1/2)$. 
 Table I collects the results for $^{133}$Cs and $^{221}$Fr. 
 
For a precise $E_1^{pv}$ measurement, it is important to suppress the large $M_1$ contribution to the
shift. To this end one may use forward-backward passages of the dressing beam which change $\hat
k$ into
$-\hat k$ and preserve the photon angular momentum.  
A systematic error may arise from  
 the mirror birefringences, but it is efficiently reduced by making   
$\pi/2$ rotations of the substrates around their axis \cite{bou74}. The $M_1$ contribution, $\propto
sign(\delta_{_F} m_{_F}) \hat E \wedge \hat k \cdot \hat B$, accessible with a single path of the beam, is a very
interesting quantity in its own right. It deserves to be measured precisely because of its exceptional sensitivity to
the accuracy of the relativistic description of the atomic system
\cite{joh99}. Here the $\delta_{_F}$-odd signature would allow discrimination against the $M_1$-allowed linear
Stark-shift arising from the far-off-resonant transitions \cite{rom99}.    

A crucial step is the calibration procedure allowing for a determination
of $E_1^{pv}$ (and $M_1$) free of uncertainties and drifts of laser power and beam geometry  
with respect to the atomic sample. To this end, one can perform  
simultaneous measurements of the E-field induced PC light-shift, namely $\Delta
\nu^{ls}_{hf}=\vert\Omega_{ind}^{\alpha}\vert^2  /2\pi \delta_{_F} $ for $\delta_{_F} \ll \Delta W /\hbar$    
amounting to $\sim 3$~Hz for Cs and a dressing intensity of 10 kW/cm$^2$. In order to discriminate $\Delta
\nu^{ls}_{hf}$ against other  contributions arising from far-off-resonant allowed transitions ($\sim -15$ Hz),  that
are nearly independent of $\omega$, one can offset the laser detuning sequentially by a small fractional amount
$\pm
\eta$ and extract the $\eta$-odd contribution.      
The ratio $\Delta \nu^{pv}_{hf}/ \eta \Delta \nu ^{ls}_{hf}$ leads directly the amplitude ratio 
$\rho^{pv}=E_1^{pv}/\alpha E$  and similarly 
$\rho^M= M_1/\alpha E$, for a single passage of the beam. From $\rho^{pv}/\rho^M $ one obtains 
$E_1^{pv}/ M_1$.  

Among all the light shifts appearing in Eq (2), we note the absence of crossed term of the type $
\Omega_{ind}^{\alpha} \Omega_{ind}^{\beta}$. Indeed, in the geometry considered here,  the associated matrix
element,
$ 2 \Omega_{ind}^{\alpha} \Omega_{ind}^{\beta} \bra F,m_{_F}\vert \vec \sigma\cdot \xi \hat  k \vert
F,m'_{_F}\ket  /2 \pi \delta_{_F}$, turns out to be off-diagonal in the $F, m_{_F}$ basis: therefore its contribution
has the same effect as a small transverse magnetic field $B^{ls}$ applied along the laser beam. In a Stark field $E$  of
100 V/cm,
$B^{ls}=2 \vert\frac {\beta}{\alpha}\vert  {\Omega_{ind}^{\alpha}}^2 / (\vert \delta_{_F} \vert  \, \mu_B)
\simeq~0.4 ~\mu$G. However, there are additional contributions associated with the vector Stark shifts coming
from the far-detuned allowed transitions such as 
$nS_{1/2}-nP_j$ (with $j=1/2, 3/2$). This is the dominant source of fictitious
magnetic field, amounting to $\sim $ 6 mG for Cs (20 mG for Fr).  Therefore, the
condition for the applied field remaining the quantization axis during the
action of the dressing beam is $B \gtrsim 10 ~$mG for Cs (30 mG for Fr).

Up to now, we have omitted from the effective dipole operator, 
Eq. (1), the PV nuclear spin-dependent terms: $\vD^{ef}_I = -i \rm{Im} E_1^{pv} (\eta \vI/I + \eta ^{\prime}  \vec
\sigma
\wedge \vI /I) $. For detunings small with respect to a $\Delta F = 0$ transition, the second operator $\propto
\eta ^{\prime} $ does not participate. The $I$-dependent contribution to $\delta E_{F,m_F}$ coming from the first
term is easily evaluated by direct transposition 
 to $\vI \cdot \ve(\xi)$ of the reasoning followed for the operator $\vec \sigma \cdot \ve(\xi)$, leading to Eq. (3).
The result involves the matrix element: $\bra F,m_{_F}\vert \hat  E \wedge \hat k \cdot  
  \vec I \vert F,m_{_F}\ket = \chi \,  (1-g_{_F}) m_{_F}$. With $\eta =
\eta^{\prime}/2= -2.1
\times 10^{-2}$ for Cs, this corresponds to a correction of $\sim 1\%$ taking 
opposite signs for the two $\Delta F = 0$ transitions. Since the weak charge contribution grows with the atomic
number much faster than that arising from the nuclear spin \cite{fla04}, we expect this 
 fractional correction not to exceed a few thousandth for the undeformed Fr isotopes, such as
$^{210}$Fr, I=6 and $^{212}$Fr, I=5, having small electric quadrupole moments \cite{isold}. In
deformed nuclei, the magnitude of the anapole moment, hence 
$\eta$, may be enhanced by factors still unpredicted. Note that for a  
dressing-beam detuned from an allowed transition, conversely, the anapole moment contribution \cite{mab07} 
dominates, the $Q_W$ one becoming less than $1\%$.
  
One can obtain more information by tuning the dressing beam close  to one
of the $\Delta F =\pm 1$ forbidden lines,  
the sole to posses an additional $M_1$ amplitude arising from the hf interaction 
\cite{bou74}. It can be identified  thanks to its contributions of opposite signs on the two
lines. It has been the subject of precise theoretical calculations and
yields an absolute reference with which to calibrate all other amplitudes from the measured ratios \cite{bou88}. 
 
Determination of $Q_W$ from the linear Stark shift in dressed $7S$ Fr atoms involves
special requirements: 
i) a  small (.01 to 1)mm$^3$ sample of $10^4$ atoms tightly
confined in an optical dipole trap, far-blue-detuned and linearly polarized to avoid large frequency shifts; ii) 
excitation and detection of a $\Delta m_{_F}=0, (m_{_F} =I-1/2) $ hf, or a Zeeman transition; iii) application of the
dressing circularly-polarized standing-wave of 10~kW/cm$^2$, and modulated detuning.
The crossed fields $E\sim$100 V/cm, $B \gtrsim$ 30 mG, are of moderate magnitudes. High
stability of B is necessary to prevent its fluctuations $\delta B$, from adding noise. 

The production of cold clouds of
$10^4$  Fr atoms inside a MOT has been reported by several groups \cite{franc}. Its fast
(non-adiabatic) transport, with optical tweezers, can be optimized to deliver an ultra-cold sample on demand to a
trap appropriately-configured at distance  \cite{dgo07}.  
There is one reported measurement of an alkali hf transition in an optical dipole trap  
\cite{chu95}, which almost matches the required conditions. 
It used the Ramsey method, with two time-separated
$\pi/2$ pulses of the microwave field. A coherence time of 4~s was observed. However, our main requirement is
stability over a few minutes. The effect of long term drifts should be reduced by fast   reversals of 
$\xi$ and $E$, thanks to the clear-cut signature of the PV shift. More recently, progress in optical dipole
traps have been reported \cite{gri00} as well as a project for   a three-dimensional optical lattice aiming at another
fundamental precision measurement, the Cs electric dipole moment (EDM) \cite{chu01}.  According to those 
authors' analysis, this design looks well adapted to 
an accurate Ramsey-type measurement of either $\Delta \nu^{pv}_{hf}$ or even 4$\times\Delta \nu^{pv}_{Ze}$.
A Raman transition with two $\sigma^{\pm}$ photons can be used to prepare the F=3, $ m_{_F}=\pm 2$ atomic
coherence  to be detected after its evolution in the dressing beam. This also makes it realistic to assume
that the projection noise limit
\cite{san99} lies within reach. The standard deviation of the frequency fluctuations, then, would be $\delta \nu= 
1/(2\pi \tau_i\sqrt{N_{at}}\,)\sqrt { T_c/\tau} $,
$\tau$ being the total measurement time. If one chooses the duration of a measurement cycle $T_c \approx
2\tau_i$, both $\delta \nu$ and $\Delta \nu ^{pv} $ (Eq. (5)) depend on $\tau_i$ in an identical manner and the
signal to noise expression $\Delta\nu^{pv}_{hf}/\delta
\nu=\Omega^{pv} \sqrt {\tau /2\Gamma_{n'S}} \sqrt{N_{at}}$ becomes independent of $\tau_i$. For an atom
number $N_{at} =10^4$, we find  $\delta\nu$(mHz)= 2.2~$\tau^{-1/2}$. Thus, assuming $\delta B < h \delta
\nu/\mu_B= $2~nG over~$\tau_i$, this gives S/N for $\Delta\nu^{pv}_{hf}$ and 4$\times\Delta
\nu^{pv}_{Ze}$ equal to 2.5 and 5, respectively, over a one hour measurement time.  
 
 In short, we have proposed a new method to measure the PV electric-dipole $nS-n'S$ forbidden transition 
amplitude, yielding the weak nuclear charge. It involves a very specific right-left asymmetry of the light-shift
induced by a dressing beam, off-resonant for the $nS-n'S$ transition, instead of the more usual asymmetry in the
transition rate. 
Since the transition is forbidden the spontaneous scattering rate can be made acceptable even at small detunings,
with moderate magnitudes of the Stark-field assisting the transition.  
This  approach avoids two-photon ionization, 
a problem encountered in previous Cs measurements  
\cite{ben99} and can be applied to a sample of only 
$\sim 10^4$ Fr atoms.  Given  the  continuous production rate of
$10^4$ cold Fr atoms/s, already achieved by several groups~\cite{franc}, measurement of the chiral PV  
Stark-shift in francium appears within reach.

We thank C. Bouchiat and M. D. Plimmer for careful reading of the manuscript. 
Laboratoire Kastler Brossel is Unit\'e de Recherche de l'Ecole Normale Sup\'erieure et de l'Universit\'e Pierre et
Marie Curie, associ\'ee au CNRS.


\vspace{-3mm} 
\begin{thebibliography}{60}
\vspace{-5mm}

 \bibitem{bou74}
M. A. Bouchiat and C. Bouchiat,  J. Phys. France, {\bf
35}, 899-927 (1974) and 
Rep. Prog. Phys.  {\bf 60},  1351  (1997). 

\bibitem{ben99}
S.~C. Bennett and C.~E. Wieman, Phys. Rev. Lett.  {\bf 82},  2484-2487 (1999). 

\bibitem{bou88}
M.-A. Bouchiat and J. Gu\'ena, {\it J. Phys. France} {\bf 49},  2037  (1988); 
C. Bouchiat and C.~A. Piketty, {\it J. Phys. France} {\bf 49},  1851 (1988).

\bibitem{fla04}
J.S.M. Ginges and V.V. Flambaum, Phys. Rep. {\bf 397}, 63 (2004) and references therein. 


\bibitem{ZAPV}
C. Bouchiat and P. Fayet, Phys. Lett. B {\bf 608}, 87 (2005); 
R. D. Young {\it et al.}, Phys. Rev. Lett {\bf 99}, 122003 (2007). 

\bibitem{gue03} 
J. Gu\'ena, {\it et al.}, Phys. Rev. Lett {\bf 90}, 143001 (2003) and Phys. Rev.
A {\bf 71}, 042108 (2005). 

\bibitem{bou05}
J. Gu\'ena, M. Lintz and M.A. Bouchiat, J. of Am. Opt. Soc. B {\bf 21}, 22 (2005).

\bibitem{franc} 
E. Gomez, L. A. Orozco, G. D. Sprouse, Rep. Prog. Phys. {\bf 69}, 79 (2006); S. Sanguinetti, private communication.  

\bibitem {san05}
S. Sanguinetti, {\it et al.}, Eur. Phys. J. D {\bf 25}, 3 (2003).

 \bibitem{lov}
C. E. Loving and P.G.H. Sandars,  
J. Phys. B {\bf 10}, 2755 (1977).

\bibitem{for93} 
E. N. Fortson, Phys. Rev. Lett. {\bf 70}, 2383 (1993).

\bibitem{joh99}
I. M. Savukov, {\it et al.} Phys. Rev. Lett. {\bf 83}, 2914 (1999).

\bibitem{mab07}
M. A. Bouchiat, Phys.Rev. Lett. {\bf 98}, 043003 (2007).

\bibitem{beb}
H.B. Bebb, Phys. Rev. A {\bf 149}, 25 (1966).

\bibitem{isold}
A. Coc {\it et al.}, Nuclear Physics, {\bf A468}, 1 (1987).

\bibitem{rom99}
M. Romalis, N. Fortson, Phys. Rev. A, {\bf 59} 4547 (1999).

\bibitem{dgo07}
A. Couvert, {\it et al.}, hal-00168928 (2007) 

\bibitem{chu95}
N. Davidson, {\it et al.}, Phys.Rev. Lett. {\bf 74}, 1311 (1995).
 
\bibitem{gri00}
R. Grimm, M. Weidemuller, and Y. B. Ovchinnikov,  Adv in At. Mol. and Opt. Phys., {\bf 42},
95 (2000).
 
\bibitem{chu01}
C. Chin {\it et al.}, Phys. Rev. A,  {\bf 63}, 033401 (2001).
A 40-fold improvement in the electron's EDM limit over that 
obtained  by E. D. Commins {\it et al.} in Tl, in 2002, requires a Stark shift detection of $0.1~\mu$Hz 
in Cs at $E$= 100 kV/cm.  

\bibitem{san99}
G. Santarelli, {\it et al.}, Phys. Rev. Lett. {\bf 82}, 4619 (1999).

 
  \end{thebibliography}
 \end{document}